\documentclass[conference]{IEEEtran}
\usepackage{cite}
\ifCLASSINFOpdf
  \usepackage[pdftex]{graphicx}
\else
\fi
\usepackage{amsmath}
\usepackage{amssymb}
\begin{document}
\title{Towards Affordable On-track Testing for Autonomous Vehicle - A Kriging-based Statistical Approach}
% Accelerate Evaluation for Lane Change Scenario Using Kriging with Optimal Sampling Scheme

\author{\IEEEauthorblockN{Zhiyuan Huang}
\IEEEauthorblockA{Department of Industrial and\\Operations Engineering\\
University of Michigan\\}
\and
\IEEEauthorblockN{Henry Lam}
\IEEEauthorblockA{Department of Industrial Engineering\\
and Operations Research\\
Columbia University\\}
\and
\IEEEauthorblockN{Ding Zhao}
\IEEEauthorblockA{Department of Mechanical Engineering\\
University of Michigan\\
Corresponding author: zhaoding@umich.edu}
}


%Telephone: (800) 555--1212, Fax: (888) 555--1212}


\maketitle

\begin{abstract}
This paper discusses the use of Kriging model in Automated Vehicle evaluation. We explore how a Kriging model can help reduce the number of experiments or simulations in the Accelerated Evaluation procedure. We also propose an adaptive sampling scheme for selecting samples to construct the Kriging model. Application examples in the lane change scenario are presented to illustrate the proposed methods.
\end{abstract}


% \fi
\IEEEpeerreviewmaketitle

\section{Introduction}

Currently, many Automated Vehicle companies adopt the Naturalistic Field Operational Test (N-FOT) \cite{FESTA-Consortium2008} approach for safety evaluation. However, this approach is inefficient because safety critical scenarios occur rarely. The required driving miles of this approach makes the testing period long, which is undesired in the competitive market.  

On the regulation side, it is the duty of NHTSA to guarantee the safety of vehicles in the market. In order to do that, they need to conduct experiments with new vehicle models \cite{Peng2012EvaluationVehicles}. However, for automated vehicles, it is hard to test its intelligence and safety because of the vast number of scenarios they meet in the daily driving. Recent research \cite{Zhao2016AcceleratedTechniques,Kim2016ImprovingMethod} have proposed methods to reduce the number of experiments required, but the amount is still infeasible for on-track tests.

In order to solve this issue and provide a safer transportation system, we propose using Kriging model \cite{ErikssonDesignExperiments} to predict the response of a real experiment. The prediction can replace the experiment effectively with an appropriate design of experiments. We present different uses of the Kriging model predictions in the Accelerated Evaluation procedure, which is proposed for Automated Vehicle tests in our previous work \cite{Zhao2016AcceleratedTechniques}. The procedure extracts and models risk events in the naturalistic driving environment for Automated Vehicle. The behavior of surrounding human-controlled vehicles is described by stochastic models. Scenarios are generated from the stochastic models and simulations or experiments are conduct to test the scenarios. The evaluation is based on the results of these independently generated and implemented tests. Since the evaluation needs an amount of tests, Importance Sampling \cite{Asmussen2007StochasticAnalysis} is used to reduce the number of required tests. We use the proposed methods to study the lane change scenario, which has been studied in \cite{Zhao2016AcceleratedTechniques,Huang2016UsingScenario}.

Besides the discussion on the use of Kriging model, we also propose schemes to select design points for Kriging model construction. These schemes can help us smartly select design points and therefore avoid doing unnecessary experiments in the model constructing procedure, while provides a better model for the prediction.

Section \ref{sec:kriging} reviews the Kriging method. We present how to use Kriging model in probability evaluation in Section \ref{sec:prob_est} and show the extension of this idea to the Accelerated Evaluation in Section \ref{sec:AE}. The optimal sampling scheme for Kriging model is in Section \ref{sec:sample}. We review the lane change scenario in Section \ref{sec:lane} and the studies of lane change scenario using these methods are presented in Section \ref{sec:exp}. Section \ref{sec:conclusion} concludes this paper.

\section{Kriging}\label{sec:kriging}
Kriging was originally developed in geostatistics and further developed by mathematicians \cite{Kleijnen2009KrigingReview}. It is a meta-modeling method that is widely used in simulation analysis. The idea is to consider the response surface as a realization of a Gaussian Random Field \cite{Rasmussen2004GaussianLearning,Staum2009BetterKriging}. 

We denote a Gaussian Random Field $y(x)$ for $x\in \mathbb{R}^d$ with mean function $\mu(x)$ and covariance function $\sigma^2(x,x')$ as \begin{equation}
	y \sim GRF(\mu,\sigma^2).
\end{equation}

For any $x\in \mathcal X$, $y(x)$ is Gaussian random variable with mean $\mu(x)$ and variance $\sigma^2(x,x)$. For $x,x' \in \mathbb{R}^d $, the covariance between $y(x)$ and $y(x')$ is $\sigma^2(x,x')$. 

For Kriging, we assume $\mu(x)=b(x) \beta$ and $\sigma^2(x,x')=\tau^2 r(x-x';\theta)$. The covariance function indicates that the variance is stationary over $x$ and the correlation function $r(\cdot)$ only depends on the value of $x-x'$.

In this paper, we use $\mu(x)=\beta$, $\beta \in \mathbb{R}$, and $r(x-x';\theta)= \prod_{i=1}^d \exp\{\theta(x_i-x_i')^2\} $, $x_i$ is the $i$th element in $x$. We use these assumptions in the following description.

Let $X$ denotes the observation matrix which is consist of $\{x^1,...,x^n\}$ and $Y$ denotes the response vector which contains $\{y^1,...,y^n\}$, where $Y \in \mathbb{R}^{n \times 1}$. We use $X$ to construct a matrix $\Sigma$, where $\Sigma_{ij}=\sigma^2(x^i,x^j)$. And let $R=\Sigma/\tau^2$. Note that $R_{ij}=r(x^i-x^j;\theta)$.

We assume that parameters $\beta$, $\tau^2$ and $\theta$ are known. Given observations $(X,Y)$, for any $x\in \mathbb{R}^d$ we have the Kriging prediction \begin{equation}
	E(y(x)|X,Y)= \beta +r(x)'R^{-1}(Y-\beta)
	\label{eq:kriging_E}
\end{equation}
and
\begin{equation}
	Var(y(x)|X,Y)=\tau^2 (1- r(x)'R^{-1}r(x)),
	\label{eq:kriging_var}
\end{equation}
where $r(x)$ returns a vector with $r(x,x^i)$ as the $i$th element.

We denote  $\mu(x|X,Y)=E(y(x)|X,Y)$ and  $\sigma^2(x|X,Y)=Var(y(x)|X,Y)$ for simplification.

We can use maximum likelihood estimation to obtain parameters $\beta$, $\tau^2$ and $\theta$ from data. We have \begin{equation}
\hat{\beta}= \frac{\sum_{i=1}^{n}y_i}{n},
\end{equation}
 and we can maximize the log likelihood function \begin{equation}
	l(\tau^2,\theta)=-\frac{1}{2} \left( n \log(2\pi) +\log(|\Sigma|) + (Y-\beta)' \Sigma^{-1}  (Y-\beta)   \right)
 \end{equation}
for $\hat{\tau}^2$ and $\hat{\theta}$.

\section{Event Probability Estimation Using Kriging}\label{sec:prob_est}

In the automated vehicle evaluation, we are interested in the probability of a type of events. Here, we discuss the use of Kriging in the probability estimation under general setting.

We use $\varepsilon$ to denote the set of events of interest. $x$ is the variable vector and the event indicator function $I_\varepsilon(x)$ returns 1 if the variables $x$ is in the event set; 0 otherwise. To evaluate the probability of events of interest, we denote $F(x)$ to be the probability distribution for $x$ and estimate the probability \begin{equation}
P(x\in \varepsilon)=E(I_\varepsilon(x))=\int I_\varepsilon(x) dF(x).
\label{eq:obj_prob}
\end{equation}

Given data set $(X,Y)$ and the Kriging model, (\ref{eq:kriging_E}) and (\ref{eq:kriging_var}) provide us a prediction for the response $y(x)$. To estimate the probability $P(x \in \varepsilon)$, we need to find a threshold $\gamma$ to construct a estimation indicator function $I_{\gamma}(y)=I\{y \geq \gamma \}$. There are two approaches to obtain an estimation for $P(x \in \varepsilon)$ using the Kriging model.

Firstly, we can take $\mu(x|X,Y)$ as the interpolation value. In this case, we can estimate $P(x \in \varepsilon)$ by \begin{equation}
\hat{P}(x \in \varepsilon) = \int I_{\gamma}(\mu(x|X,Y)) dF(x).
\label{eq:ke_simple}
\end{equation}
This approach simplifies the inference from the Kriging model and is easy to implement. We note that the integral in (\ref{eq:ke_simple}) is generally hard to compute analytically. We can use sample estimation for the integral.

To fully exploit the Kriging model, we can estimate $P(x \in \varepsilon)$ by \begin{equation}
\hat{P}(x \in \varepsilon) = \int E\left[I_{\gamma}(y(x)) \right]  dF(x).
\label{eq:ke_complex}
\end{equation}
The integrand is the expectation of $I_{\gamma}(y(x))$ using the Kriging prediction on $x$. We can further expand the integrand as\begin{equation}
E\left[I_{\gamma}(y(x)) \right]=P(y(x)\geq \gamma)=1-\Phi(\frac{\gamma-\mu(x|X,Y)}{\sqrt{\sigma^2(x|X,Y)}}),
\end{equation}
where $\Phi(\cdot)$ denote the cumulative density function of standard Gaussian distribution. Therefore, we can write (\ref{eq:ke_complex}) as \begin{equation}
\hat{P}(x \in \varepsilon) = 1- \int \Phi(\frac{\gamma-\mu(x|X,Y)}{\sqrt{\sigma^2(x|X,Y)}})  dF(x).
\end{equation}
This integral also requires sample estimation in practice.

\section{Improving the Accelerated Evaluation Using Kriging} \label{sec:AE}

\subsection{Accelerated Evaluation}

Accelerated Evaluation concept is proposed in our previous work \cite{Zhao2016AcceleratedTechniques}. The procedure focus on evaluating Automated Vehicle under interaction with human-controlled vehicles. The basic idea is to model human-controlled vehicles as stochastic models and to conduct simulation to test random generated scenarios. The probability of event of interest can be estimated from the simulations. Since we are generally interested in rare events, we use variance reduction techniques to accelerate the evaluating procedure. The evaluation procedure contains four steps \cite{Huang2016UsingScenario}:
\begin{enumerate}
	\item Model the behaviors of the “primary other vehicles” (POVs) represented by $f(x)$ as the major disturbance to the AV using large-scale naturalistic driving data \label{enum:step1}
	\item Skew the disturbance statistics from $f(x)$ to modified statistics $f^{*}(x)$ (accelerated distribution) to generate more frequent and intense interactions between AVs and POVs \label{enum:step2}
	\item Conduct “accelerated tests” with $f^{*}(x)$ \label{enum:step3}
	\item Use the Importance Sampling (IS) theory to “skew back” the results to understand real-world behavior and safety benefits \label{enum:step4}
\end{enumerate}

The Step \ref{enum:step2} and Step \ref{enum:step4} involves the rare event indicator function $I_\varepsilon(\cdot)$, where we can replace it by the Kriging estimation indicator function. We discuss the two steps separately in the following subsections.

\subsection{Improving the Importance Sampling Procedure}

The Importance Sampling method is a variance reduction technique for Monte Carlo simulation. We can write (\ref{eq:obj_prob}) as

\begin{equation}
	E[I_\varepsilon(X)]=\int I_\varepsilon(x) dF = \int I_\varepsilon(x) \frac{dF}{dF^*} dF^*.
\end{equation}
This indicates that the expectation of interest equals to $E[I_\varepsilon(X) \frac{dF}{dF^*}]$ over distribution $F^*$. Therefore, instead of sampling from $F$ and compute the sample mean of $I_\varepsilon(X)$, we can sample from $F^*$ and compute the sample mean of $I_\varepsilon(X) \frac{dF}{dF^*}$. $F^*$ is called Importance Sampling distribution and a good selection of $F^*$ provides very efficient estimator in the sense of the variance of the estimator. 

Suppose that $F^*$ is given, the Importance Sampling estimator for $P(x\in \varepsilon)$ is \begin{equation}
\hat{P}(x \in \varepsilon) = \frac{1}{N} \sum_{i=1}^{N} I_\varepsilon(x_i) \frac{dF}{dF^*},
\label{eq:is_estimator}
\end{equation}
where $x_i$'s are generated from the Importance Sampling distribution $F^*$. Similar to Section \ref{sec:prob_est}, we have two ways to replace the original indicator function $I_\varepsilon(x_i)$ by the information from the Kriging model.

To replace $I_\varepsilon(x_i)$ by $ I_{\gamma}(\mu(x_i|X,Y))$, we have
\begin{equation}
\hat{P}(x \in \varepsilon) =  \frac{1}{N} \sum_{i=1}^{N} I_{\gamma}(\mu(x_i|X,Y)) \frac{dF}{dF^*}.
\label{eq:is_simple}
\end{equation}

To replace $I_\varepsilon(x_i)$ by $ E\left[I_{\gamma}(y(x_i)) \right]$, we have
\begin{equation}
\hat{P}(x \in \varepsilon) = \frac{1}{N} \sum_{i=1}^{N} E\left[I_{\gamma}(y(x_i)) \right] \frac{dF}{dF^*}.
\label{eq:is_complex}
\end{equation}

We note that to obtain a valid estimation of the probability $P(x \in \varepsilon)$, the ideal case is that the replacement of $I_\varepsilon(x_i)$ returns the same output as the original function. Therefore, when we have an accurate Kriging model, (\ref{eq:is_simple}) is a better choice, consider that (\ref{eq:is_complex}) gives a more vague prediction of the value of $I_\varepsilon(x_i)$.

Since the estimation of $P(x \in \varepsilon)$ is the final output of the Accelerated Evaluation procedure, we recommend to use the real indicator function in the estimation, unless the cost of the simulation is not affordable.
\subsection{Improving the Cross Entropy Procedure}

We proposed in \cite{Zhao2016AcceleratedTechniques,Huang2016UsingScenario} to use Cross Entropy method \cite{Kroese2013a} in the Step \ref{enum:step2}, where the rare event indicator function is involved throughout the procedure.

The Cross Entropy is a method to construct effective Importance Sampling distributions. The key idea of Cross Entropy method is to optimize the parameter $\theta$ in a parametric distribution family $f_\theta(\cdot)$. The optimal parameter provides a distribution that is most similar to the optimal Importance Sampling distribution \begin{equation}
	f^{**}(x)=\frac{I_\varepsilon(x)f(x)}{{P}(x \in \varepsilon)} .
\end{equation}
regards to the Kullback-Leibler distance between them. The Cross Entropy method iterates for the optimal parameter and in each iteration an optimization problem is solved. The objective function of the optimization problem is \begin{equation}
	\max_\theta \ \frac{1}{N} \sum_{i=1}^{N} I_\varepsilon(x_i) \frac{f(x_i)}{f_{\theta_s}(x_i)} \ln f_\theta(x_i),
    \label{eq:ce_obj}
\end{equation}
where $x_i$'s are sampled from distribution $f_{\theta_s}$ and $\theta_s$ is given.

Again, given the Kriging model, we can replace the original indicator function $I_\varepsilon(x_i)$ by $ I_{\gamma}(\mu(x_i|X,Y))$ or $ E\left[I_{\gamma}(y(x_i)) \right]$. Since the optimality of the parameter only affects the efficiency of the Importance Sampling procedure, the accuracy of the solution of (\ref{eq:ce_obj}) is not as crucial as the estimation of $P(x \in \varepsilon)$. Either replacement is applicable under this situation.

We note that the simulations for $I_\varepsilon(x_i)$ are not as necessary as those in the Importance Sampling part and the results of $I_\varepsilon(x_i)$ cannot be used in later steps. Therefore, when the cost of the simulation for $I_\varepsilon(x_i)$ is expensive, we do not want to spend too much efforts on the Cross Entropy procedure. In this case, the Kriging model can provide a large saving on almost no cost. Comparing with using Kriging model information in the Importance Sampling procedure, replacing the real simulation with Kriging Model estimation is more recommended.

\section{Optimal Sampling Scheme For Constructing Kriging Model} \label{sec:sample}

To construct a Kriging model, we need to have sample set $(X,Y)$ to predict the response. Adding new samples to the set can improve the prediction. Instead of picking up an arbitrary sample, we want to smartly select at what $x$ to sample. Here we present two approaches to guide the selection of sample points.

We assume that we have a sample set $(X_n,Y_n)$ with $n$ samples and we are looking for a new sample $x$ from the selection set $\mathcal{X}$. Note that $\mathcal{X}$ can be continuous or discrete.

\subsection{Point Optimal Sampling Scheme}

We first propose an intuitive scheme for selecting sample points: we sample at the point that the prediction model cannot provide much information. Using this idea, we can check among $\mathcal{X}$ to find out at which $x \in \mathcal{X}$, the estimated return is the most ``ambiguous''. Since the scheme focuses on the return at each point $x$, we refer it as point optimal sampling scheme.

In Section \ref{sec:prob_est}, we proposed to estimate $I_\varepsilon(x_i)$ from the Kriging model using $ I_{\gamma}(\mu(x_i|X_n,Y_n))$, the response of $ I_{\gamma}(\mu(x_i|X_n,Y_n))$ is our main concern regards to the Kriging model. We can use the inference from Kriging model to find which $x \in \mathcal{X}$ has the largest probability to have a different outcome after sampling, which we define as the most ambiguous point.

If $\mu(x_i|X_n,Y_n) \geq \gamma$, the probability of sampling a different outcome is 
\begin{equation}
	P(y(x) < \gamma)=\Phi(\frac{\gamma-\mu(x|X_n,Y_n)}{\sqrt{\sigma^2(x|X_n,Y_n)}});\label{eq:point_p_left}
\end{equation}
otherwise, when  $\mu(x_i|X_n,Y_n) < \gamma$, the probability is
\begin{equation}
P(y(x) \geq \gamma)=\Phi(\frac{-(\gamma-\mu(x|X_n,Y_n))}{\sqrt{\sigma^2(x|X_n,Y_n)}}).\label{eq:point_p_right}
\end{equation}

Based on the probability in (\ref{eq:point_p_left}) and (\ref{eq:point_p_right}), we want to sample from $x \in \mathcal{X}$ has the largest probability to obtain a sample with a different outcome from the Kriging prediction. Therefore, we sample at $x$, such that \begin{equation}
	x= \arg \max_{x \in \mathcal{X}} \Phi(\frac{-|\gamma-\mu(x|X_n,Y_n)|}{\sqrt{\sigma^2(x|X_n,Y_n)}}).
    \label{eq:pnt1}
\end{equation}

This scheme searches for the most ambiguous point in the potential sampling set to improve the Kriging model. It is easy to implement, but it only consider the outcome of the indicator function $ I_{\gamma}(\mu(x_i|X_n,Y_n))$. 

In the cases we mentioned in Sections \ref{sec:prob_est} and \ref{sec:AE}, when we use $ E\left[I_{\gamma}(y(x_i)) \right]$ to replace the original indicator function, this scheme does not work any more. Instead of checking the probability get a different outcome, we consider how much difference a new sample at $x$ would bring to the value of $ E\left[I_{\gamma}(y(x_i)) \right]$. This can be measured by the variance of $I_{\gamma}(y(x_i))$. Therefore, in this case, we only need to pick up $x$ with largest $Var \left[I_{\gamma}(y(x_i))\right]$, which writes as\begin{equation}
	x = \arg \max_{x \in \mathcal{X}}E\left[I_{\gamma}(y(x_i)) \right] -E\left[I_{\gamma}(y(x_i)) \right]^2.
    \label{eq:pnt2}
\end{equation}

\subsection{Objective Optimal Sampling Scheme}

Instead of focusing on the difference a sample might make on a point, we can also select the next sample point by checking how much improvement it can contribute for the objective of interest. Here, we use the event probability estimation in Section \ref{sec:prob_est} as examples.

We firstly introduce some notations. $E_n[\cdot]$ represents the expectation over $y(x)$ with distribution from the Kriging model, i.e., Gaussian distribution with mean $\mu(x|X_n,Y_n)$ and variance $\sigma^2(x|X_n,Y_n)$. We use $(X_{n+1},Y_{n+1})$ to denote the original sample set $(X_n,Y_n)$ added by a new sample $(x,Y(x))$, where $Y(x)$ is a realization of random variable $y(x)$. $E_{n+1}[\cdot]$ represents the corresponding expectation over $y(x)$ with mean $\mu(x|X_{n+1},Y_{n+1})$ and variance $\sigma^2(x|X_{n+1},Y_{n+1})$.

To simplify the description, we use $P_n$ to represent the probability estimation with the existing sample set $(X_n,Y_n)$, $P_{n+1}$ to represent the probability estimation with $(X_{n+1},Y_{n+1})$. We want to find select new sample point $x$ as\begin{equation}
	x = \arg \max_{x \in \mathcal{X}} E_n \left[ (P_n-P_{n+1})^2  \right].
    \label{obj}
\end{equation}
Note that the expectation is over the realization $Y(x)$ in the new sample set $(X_{n+1},Y_{n+1})$.

Consider the estimation (\ref{eq:ke_simple}), we have \begin{equation}
	P_n=\int I_{\gamma}(\mu(\omega|X_n,Y_n))  dF(\omega)
    \label{obj1_1}
\end{equation}
and 
\begin{equation}
P_{n+1}=\int I_{\gamma}(\mu(\omega|X_{n+1},Y_{n+1}))  dF(\omega).
\label{obj1_2}
\end{equation}

For the estimation (\ref{eq:ke_complex}), we have \begin{equation}
	P_n=\int E_n\left[I_{\gamma}(y(\omega)) \right] dF(\omega)
    \label{obj2_1}
\end{equation}
and 
\begin{equation}
P_{n+1}=\int E_{n+1}\left[I_{\gamma}(y(\omega)) \right] dF(\omega).
\label{obj2_2}
\end{equation}

By replacing $P_{n}$ and $P_{n+1}$ by other estimation objective, we can apply this idea to other cases, such as the cases described in Section \ref{sec:AE}.

\subsection{Discussion on The Sampling Schemes}

Firstly, we note that these two types of sampling scheme can be used to form an adaptive sampling procedure for the Kriging model. The adaptive sampling procedure is suggested to be the following:
\begin{enumerate}
\item Start with arbitrary existing sample set $(X_n,Y_n)$.
\item Generate a sample selection set $\mathcal{X}$. A reasonable approach to generate the selection set is to discretize the sampling space. \label{item:start}
\item Compute the sampling criterion and decide at which point to sample.
\item Add the new sample and its response into the existing sample set.\label{item:end}
\item Iterate from \ref{item:start} to \ref{item:end}.
\end{enumerate}
This procedure provides a sequential sample selection scheme for finding reasonable sampling points. 

Secondly, the point optimal scheme only focuses on the Kriging model itself, while objective optimal scheme considers the sampling points regards to the objective estimation. We suggest to use the objective optimal scheme in the Accelerated Evaluation procedure. In other cases, if the goal of constructing a Kriging model is not clear when the sample set is constructed, the objective function is not available. The point optimal scheme is feasible regardless to how we use the Kriging model. 

\section{The Lane Change Scenario}\label{sec:lane}

In this paper, we use the lane change scenario as an example to illustrate the proposed methods. The lane change scenario is the defined as when an Automated Vehicle (AV) is driving, a human-controlled vehicle driving in front of the AV start to cut into the AV's lane. In this scenario, we assume the condition for the two vehicles at the moment a lane change starts is random. With the starting condition, we can simulate the interaction of the vehicles in the lane change procedure. We extracted lane change data of naturalistic driving from the Safety Pilot Model Deployment (SPMD) database \cite{Bezzina2014}. We use these data to model the randomness of the starting condition. Three key variables can capture the effects of gap acceptance of the lane changing vehicle: velocity of the lead vehicle $v$, range to the lead vehicle $R$ and time to collision $TTC$. $TTC$ was defined as:
\begin{equation}
	TTC=- \frac{R}{\dot{R}},
\end{equation}
where $\dot{R_L}$ is the relative speed. The Automated Vehicle model we used in simulation is constructed using Adaptive Cruise Control (ACC) and Autonomous Emergency Braking (AEB) \cite{Ulsoy2012a} systems. 

Our aim is to study events (generally risk events) occurs during the lane change procedure with this model. In \cite{Zhao2016AcceleratedTechniques} and \cite{Huang2016UsingScenario}, we evaluated the probability for crash, conflict and injury as events of interest. We use $\varepsilon$ to represent the set of events of interest and use an event indicator function $I_\varepsilon(x)$ that returns $1$ (event occurs) or $0$ (safe) to represent the simulation. Here, $x$ is a vector variable that contains $v$, $R$ and $TTC$.

\begin{figure}[t]
	\centering
	\includegraphics[width=0.8\linewidth]{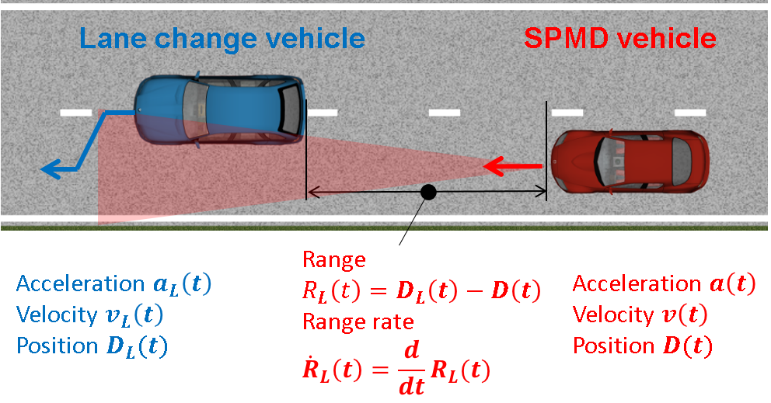}
	\caption{Lane change data collected by SPMD vehicle.}
	\label{fig:lane}
\end{figure}

\section{Analysis on The Lane Change Scenario} \label{sec:exp}

In this section, we use an example in the lane change problem to present the methods we proposed. Our objective is to check in the lane change scenario, when the velocity of the leading vehicle $v$ is low, what is the probability that the two vehicles have a minimum range smaller than 2 meters during the lane change procedure. From the study of \cite{Zhao2016AcceleratedTechniques}, we know that when the velocity $v$ is between 5 to 15 $m/s$, the other two variables $R^{-1}$ and $TTC^{-1}$ are independent to each other. $TTC^{-1}$ can be modeled by exponential distribution and $R^{-1}$ can be modeled by Pareto distribution. Here we denote $x=[TTC^{-1},R^{-1}]$ and $I_\varepsilon(x)$ for the event of interest.

Instead of using maximum likelihood estimation, we select parameters $\beta$, $\tau^2$ regarding the nature of the problem. We note the response of data are 1 or 0, where 1 represents the event of interest happened. Since 1 rarely occurs, we want $\beta=0$, so when there is no information, we assume the event would not happen. We set $\gamma=0.5$ in this case. We use $\tau^2=0.01$, since when the value of the $\mu(x|X,Y)$ is low, we want it to be 2 to 3 times standard deviation away from $\gamma$. 

\begin{figure}[t]
	\centering
	\includegraphics[width=\linewidth]{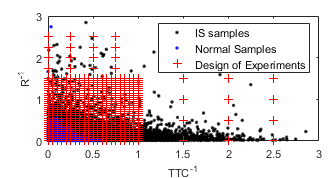}
	\caption{Design of Experiments for Kriging Modeling.}
	\label{fig:DOE}
\end{figure}

We design experiments as shown in Fig. \ref{fig:DOE}. We observed the data from original distribution and an IS distribution. The 685 design points are selected to cover high probability region. We then use 5,000 data with response to select an appropriate $\theta$ by comparing the Kriging return $ I_{\gamma}(\mu(x|X,Y))$ with $I_\varepsilon(x)$. Among the 5,000 data, 5 events of interest happened. Using $\theta=100$, we successfully predict 3 of them and using $\theta=50$ gives 4 of them. All 0 responses are correctly predict in both case. Therefore, we use these two values of $\theta$ in the following experiments. We refer $\theta=50$ as low $\theta$ and $\theta=100$ as high $\theta$.

\subsection{Event Probability Estimation}

\begin{figure}[t]
	\centering
	\includegraphics[width=\linewidth]{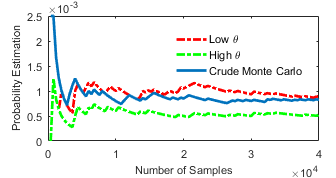}
	\caption{Comparison of Probability Estimation.}
	\label{fig:pro_est}
\end{figure}

We directly simulate for the event probability using crude Monte Carlo method. We also apply the two approaches in Section \ref{sec:prob_est} using the Kriging model we construct to estimate the probability. Fig. \ref{fig:pro_est} presents the comparison of the crude Monte Carlo approach and the proposed methods. Since we use a small $\tau^2$, (\ref{eq:ke_simple}) and (\ref{eq:ke_complex}) gives very similar results, we only present results using (\ref{eq:ke_simple}). We note that the Kriging model can provide a roughly correct estimation without doing any extra experiments.

\subsection{Improving Accelerated Evaluation}

Using Kriging model in the Cross Entropy method and the Importance Sampling method are very similar. In both cases, the indicator function is estimated and then multiplied by a score function. We only present an example on the Importance Sampling method, where the accuracy of the estimation is more important.

\begin{figure}[t]
	\centering
	\includegraphics[width=\linewidth]{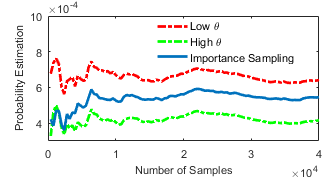}
	\caption{Comparison of Importance Sampling.}
	\label{fig:is_est}
\end{figure}

Since the two approaches (\ref{eq:is_simple}) and (\ref{eq:is_complex}) have similar performance, we only present the results using (\ref{eq:is_simple}). Fig.~\ref{fig:is_est} shows the comparison of the Importance Sampling method and the estimation using Kriging model with two different $\theta$. The estimation using Kriging is still reasonable in this case. We note that the trend of the three lines are very similar, this indicates that the prediction of the Kriging models are very stable.

\subsection{Sampling Scenario Selection}

The design of experiments for the Kriging model in the previous examples is based on observation of data. Here, we use an example problem to illustrate the optimal sample selection methods in Section \ref{sec:sample}.

\begin{figure}[t]
	\centering
	\includegraphics[width=\linewidth]{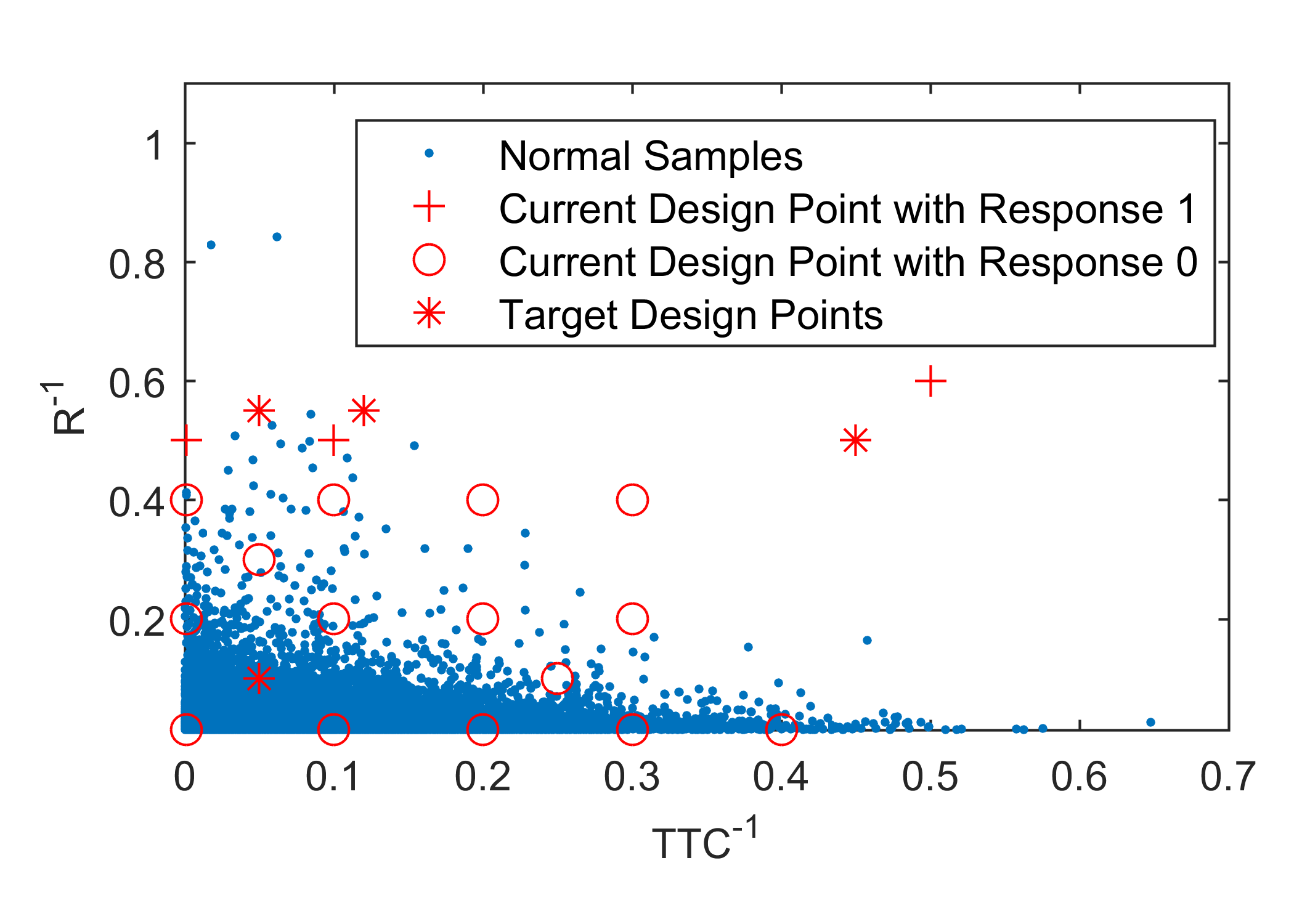}
	\caption{The setting of the sample selection example.}
	\label{fig:exp3_set}
\end{figure}

Fig. \ref{fig:exp3_set} shows the sample distributions and existing design points. We assume that we want to select a sample from the 4 target design points to improve the Kriging model. This is the key step for the adaptive sampling procedure. We note that the number of sample selections is largely reduced in this example, but the method is obviously applicable for a large number of selections.

Table \ref{table:exp3} presents the results using different approaches we proposed. The numbers in rows 3 to 6 are objective function values for the 4 different approaches. 

We note that ``Pnt 1'' refers to the function value of (\ref{eq:pnt1}), ``Pnt 2'' refers to (\ref{eq:pnt2}). These two criteria are point optimal schemes. Since points B and C are closer to some known points, the Kriging model provides more information for them than the points A and D. If we use point optimal criterion, we want to select point D, since the prediction of the Kriging model is the most likely to change.

For the objective optimal schemes, the objective function is (\ref{obj}). For $P_{n}$ and $P_{n+1}$, ``Obj~1'' uses (\ref{obj1_1}) and (\ref{obj1_2}), while ``Obj~2'' uses (\ref{obj2_1}) and (\ref{obj2_2}) respectively.
In this case, we use the event probability estimation as the estimation objective. Point D become least important according to the objective optimal criterion. This is because it is away from high probability region of the randomness. For objective optimal criterion, point A is more important. The reason is that the point has a large probability density of data samples.

We use this example to show that the difference between the two types of sampling selection schemes. In this case, the objective optimal schemes provide a more reasonable choice, consider that we want to know more information on the high probability region.

\begin{table}[t]
\centering
\caption{The results from 4 different approaches are present regards to the target design points.}
\renewcommand{\arraystretch}{2}
\label{table:exp3}
\resizebox{\columnwidth}{!}{%
\begin{tabular}{|l|l|l|l|l|}
\hline
        & A          & B           & C           & D          \\ \hline
Coord   & (0.05,0.1) & (0.12,0.55) & (0.05,0.55) & (0.45,0.5) \\ \hline
Pnt 1 & 0.298   & 0.112    & 0.0495   & 0.469   \\ \hline
Pnt 2 & 0.209   & 0.0996    & 0.0471   & 0.249   \\ \hline
Obj 1   & $2.63 \times 10^{-5}$   & $1.69 \times 10^{-7}$    & $1.75 \times 10^{-7}$    & $1.69 \times 10^{-7}$   \\ \hline
Obj 2   & 0.0082   & $4.26 \times 10^{-8}$    & $6.76 \times 10^{-8}$    & $3.13 \times 10^{-11}$   \\ \hline
\end{tabular}}
\end{table}

\section{Conclusion}\label{sec:conclusion}

This paper presents several approaches to use Kriging model in Automated Vehicle evaluating. The Kriging model provides a reasonable estimation for different objectives with no simulation cost. We present schemes to select design points for Kriging model. The objective optimal sampling schemes are suggested for the Automated Evaluation procedure.

% \section*{Acknowledgment}

% The authors acknowledge support from the University of Michigan Mobility Transformation Center: a project under grant number N021552.



%\bibliographystyle{IEEEtran}

\bibliographystyle{IEEEtran}
\bibliography{Mendeley_MTC_AE.bib}

\begin{thebibliography}{10}
\providecommand{\url}[1]{#1}
\csname url@samestyle\endcsname
\providecommand{\newblock}{\relax}
\providecommand{\bibinfo}[2]{#2}
\providecommand{\BIBentrySTDinterwordspacing}{\spaceskip=0pt\relax}
\providecommand{\BIBentryALTinterwordstretchfactor}{4}
\providecommand{\BIBentryALTinterwordspacing}{\spaceskip=\fontdimen2\font plus
\BIBentryALTinterwordstretchfactor\fontdimen3\font minus
  \fontdimen4\font\relax}
\providecommand{\BIBforeignlanguage}[2]{{%
\expandafter\ifx\csname l@#1\endcsname\relax
\typeout{** WARNING: IEEEtran.bst: No hyphenation pattern has been}%
\typeout{** loaded for the language `#1'. Using the pattern for}%
\typeout{** the default language instead.}%
\else
\language=\csname l@#1\endcsname
\fi
#2}}
\providecommand{\BIBdecl}{\relax}
\BIBdecl

\bibitem{FESTA-Consortium2008}
{FESTA-Consortium}, ``{FESTA Handbook Version 2 Deliverable T6.4 of the Field
  opErational teSt supporT Action},'' FESTA, Tech. Rep., 2008.

\bibitem{Peng2012EvaluationVehicles}
H.~Peng and D.~Leblanc, ``{Evaluation of the Performance and Safety of
  Automated Vehicles},'' \emph{White Pap. NSF Transp. CPS Work}, 2012.

\bibitem{Zhao2016AcceleratedTechniques}
D.~Zhao, H.~Lam, H.~Peng, S.~Bao, D.~J. LeBlanc, K.~Nobukawa, and C.~S. Pan,
  ``{Accelerated Evaluation of Automated Vehicles Safety in Lane-Change
  Scenarios Based on Importance Sampling Techniques},'' \emph{IEEE Transactions
  on Intelligent Transportation Systems}, 2016.

\bibitem{Kim2016ImprovingMethod}
\BIBentryALTinterwordspacing
Y.~Kim and M.~J. Kochenderfer, ``{Improving Aircraft Collision Risk Estimation
  Using the Cross-Entropy Method},'' \emph{Journal of Air Transportation},
  vol.~24, no.~2, pp. 55--61, 4 2016. [Online]. Available:
  \url{http://arc.aiaa.org/doi/10.2514/1.D0020}
\BIBentrySTDinterwordspacing

\bibitem{ErikssonDesignExperiments}
\BIBentryALTinterwordspacing
L.~Eriksson, E.~Johansson, and N.~Kettaneh-Wold, ``{Design of experiments},''
  \emph{dynacentrix.com}. [Online]. Available:
  \url{https://www.dynacentrix.com/telecharg/Modde/Livredoe.pdf}
\BIBentrySTDinterwordspacing

\bibitem{Asmussen2007StochasticAnalysis}
S.~Asmussen and P.~Glynn, \emph{{Stochastic Simulation: Algorithms and
  Analysis}}.\hskip 1em plus 0.5em minus 0.4em\relax Springer, 2007.

\bibitem{Huang2016UsingScenario}
\BIBentryALTinterwordspacing
Z.~Huang, D.~Zhao, H.~Lam, D.~J. LeBlanc, and H.~Peng, ``{Using the Piecewise
  Mixture Model to Evaluate Automated Vehicles in the Frontal Cut-in
  Scenario},'' 10 2016. [Online]. Available:
  \url{http://arxiv.org/abs/1610.09450}
\BIBentrySTDinterwordspacing

\bibitem{Kleijnen2009KrigingReview}
\BIBentryALTinterwordspacing
J.~P. Kleijnen, ``{Kriging metamodeling in simulation: A review},''
  \emph{European Journal of Operational Research}, vol. 192, no.~3, pp.
  707--716, 2 2009. [Online]. Available:
  \url{http://linkinghub.elsevier.com/retrieve/pii/S0377221707010090}
\BIBentrySTDinterwordspacing

\bibitem{Rasmussen2004GaussianLearning}
\BIBentryALTinterwordspacing
C.~E. Rasmussen, ``{Gaussian Processes in Machine Learning}.''\hskip 1em plus
  0.5em minus 0.4em\relax Springer Berlin Heidelberg, 2004, pp. 63--71.
  [Online]. Available:
  \url{http://link.springer.com/10.1007/978-3-540-28650-9{\_}4}
\BIBentrySTDinterwordspacing

\bibitem{Staum2009BetterKriging}
\BIBentryALTinterwordspacing
J.~Staum, ``{Better simulation metamodeling: The why, what, and how of
  stochastic kriging},'' in \emph{Proceedings of the 2009 Winter Simulation
  Conference (WSC)}.\hskip 1em plus 0.5em minus 0.4em\relax IEEE, 12 2009, pp.
  119--133. [Online]. Available:
  \url{http://ieeexplore.ieee.org/document/5429320/}
\BIBentrySTDinterwordspacing

\bibitem{Kroese2013a}
D.~P. Kroese, R.~Y. Rubinstein, and P.~W. Glynn, \emph{{The Cross-Entropy
  Method for Estimation}}.\hskip 1em plus 0.5em minus 0.4em\relax Elsevier
  B.V., 2013, vol.~31.

\bibitem{Bezzina2014}
D.~Bezzina and J.~R. Sayer, ``{Safety Pilot: Model Deployment Test Conductor
  Team Report},'' NHTSA, Tech. Rep., 2014.

\bibitem{Ulsoy2012a}
A.~G. Ulsoy, H.~Peng, and M.~{\c{C}}akmakci, \emph{{Automotive control
  systems}}.\hskip 1em plus 0.5em minus 0.4em\relax Cambridge University Press,
  2012.

\end{thebibliography}

\end{document}